\magnification=1200
\input harvmac 
\hfuzz=15pt 
\font\male=cmr9 
 
\font\sfont=cmbx10 scaled\magstep1


  \def\bbz{Z\!\!\!Z}
\def\bbc{C\kern-6.5pt I}  \def\bbn{I\!\!N}
\def\a{\alpha}  \def\g{\gamma} \def\d{\delta}
\def\k{\kappa} 
\def\xiL#1#2{\xi^{1\,\cdots\,#1}_{1\,\cdots\,#2}\,} 
\def\hp{\hat{\varphi}}  \def\l{\lambda} \def\ca{{\cal A}} 
\def\L{\Lambda} 
\def\cc{{\cal C}}  
\def\cu{{\cal U}}  \def\cp{{\cal P}}  
  \def\hd{{\hat{\cal D}}}  \def\hm{{\hat M}} 
\def\cz{{\cal Z}} 
\def\cd{D} 
\def\r{\rho} \def\eps{\epsilon} \def\s{\sigma} 
\def\uq{U_q(sl(3))}  \def\un{U_q(sl(n))} 
\def\ag{{{\cal A}_g}} \def\an{SL_q(n)}
 \def\x{\xi} \def\z{\zeta} 
\def\lg{\langle} \def\rg{\rangle} \def\ve{\varepsilon}
\def\vf{\varphi} \def\y{\eta} 
\def\tv{{\tilde\vf}} \def\tc{{\hat \cc}}  
\def\rra{\longrightarrow}  \def\vr{\vert}

  \def\br{{\bar r}}  
 \def\ci{{\cal I}} 
   
\def\pd{\partial}    
\def\hr{{\hat r}} 
\def\bc{{\bar c}} 


\baselineskip=12pt

ASI-TPA/10/93 (October 1993), $\quad$ hepth/9405150;

Published in: J. Phys. A: Math. Gen. {\bf 27} (1994) 4841-4857;

Note Added published in: 
J. Phys. A: Math. Gen. {\bf 27} (1994) 6633-6634.

\vskip 1.5cm

\baselineskip=16pt plus 2pt minus 1pt 

\centerline{{\sfont q - Difference Intertwining Operators for 
U$_{{\bf q}}$(sl(n))~: }} 
\centerline{{\sfont General Setting and the Case ~n=3}}
\baselineskip=12pt 

\vskip 1.5cm 

\centerline{{\bf V.K. Dobrev$^{*}$} }
\footnote{}{$^{*}$ 
~\male{Permanent address : Bulgarian Academy of Sciences,
Institute of  Nuclear Research and Nuclear Energy, 72
Tsarigradsko Chaussee, 1784 Sofia, Bulgaria.}} 

\vskip 0.5cm

\centerline{Arnold Sommerfeld Institute
for Mathematical Physics}  
\centerline{Technical University Clausthal}  
\centerline{Leibnizstr. 10, 38678 Clausthal-Zellerfeld, Germany}

\baselineskip=16pt plus 2pt minus 1pt 

\parskip=7pt plus 1pt

\vskip 2cm

\centerline{\bf Abstract}

We  construct representations ~$\hat\pi_{\br}$~ of the quantum
algebra ~$\un$~ labelled by $n-1$ complex numbers ~$r_i$~ and
acting in the space of formal power series of ~$n(n-1)/2$~
non-commuting variables.  These variables generate a flag
manifold  of the matrix  quantum group  ~$\an$~ which is dual to
~$\un$~. The conditions for reducibility of ~$\hat\pi_{\br}$~
and the procedure for the construction of the ~$q$ - difference
intertwining operators are given. The representations 
and $q$ - difference intertwining operators are given in the most
explicit form for $n=3$.  

\vfill\eject  

\baselineskip=16pt plus 2pt minus 1pt 

\parskip=7pt plus 1pt

\null\bigskip 

\noindent {\bf 1. ~~Introduction} 

Invariant differential equations ~$\ci ~f ~=~ 0$~ play a very
important role in the description of physical symmetries -
recall, e.g., the examples of Dirac, Maxwell equations, 
(for more examples cf., e.g., \ref\BR{A.O.Barut and R. R\c aczka, 
{\it Theory of Group Representations and Applications}, \hfil\break  
II edition, (Polish Sci. Publ., Warsaw, 1980).}). 
It is an important and yet unsolved problem to find such equations 
for the setting of quantum groups, where they are expected 
as $q$-difference equations, especially, in the case 
of non-commuting variables. 

The approach to this problem used here relies on the following. 
In the classical situation the invariant differential operators 
~$\ci$~ giving the equations above may be described as 
operators intertwining representations of complex and real
semisimple Lie groups \ref\KS{A.W. Knapp and E.M. Stein, Ann.
Math. {\bf 93} (1971) 489; ~Inv. Math. {\bf 60} (1980) 9.},
\ref\Ko{B. Kostant, Lecture Notes in Math., Vol. 466
(Springer-Verlag, Berlin, 1975) p.  101.}, \ref\Zh{D.P.
Zhelobenko, Math. USSR Izv. {\bf 10} (1976) 1003.}, 
\ref\Dob{V.K. Dobrev, Rep. Math. Phys. {\bf 25} (1988) 159.}. 
There are many ways to find such operators, cf., e.g., \BR, 
however, most of these rely on constructions which are not 
available for quantum groups. Here we shall apply a procedure 
\Dob\ which is rather algebraic and can be generalized almost 
straightforwardly to quantum groups. According to this procedure 
one first needs to know these constructions for the complex 
semisimple Lie groups since the consideration of a real semisimple 
Lie group involves also its complexification. That  
is why we start here with the case of ~$\un$ ~(we write 
$sl(n)$ instead of $sl(n,\bbc)$). For the procedure one  
needs ~$q$-difference realizations of the representations 
in terms of functions of non-commuting variables. 
Until now such a realization of the representations 
and of the intertwining operators was found only for  
a Lorentz quantum algebra (dual to the matrix Lorentz quantum group 
of \ref\WZ{S.L. Woronowicz and S. Zakrzewski, Quantum
Lorentz group having Gauss decomposition property, ETH Z\"urich
and TU Clausthal preprint, (July 1991).}) in 
\ref\DDF{L. Dabrowski, V.K. Dobrev and R. Floreanini, 
J. Math. Phys. {\bf 35} (1994) 971.}. 
The construction in \DDF\ (also applying 
the procedure of \Dob) involves 
two $q$-commuting variables $\y\bar\y = q\bar\y\y$ 
and uses the complexification 
~$U_q(sl(2)) \otimes U_q(sl(2))$~ of the Lorentz quantum algebra. 

In the present paper following the 
mentioned procedure we  construct representations 
~$\hat\pi_{\br}$~ of ~$\un$~ labelled by ~$n-1$~ complex numbers
~$\br ~=~ \{ r_1,\ldots, r_{n-1}\}$~ and acting in the spaces of
formal power series of ~$n(n-1)/2$~ non-commuting (for $n>2$) variables
~$Y_{ij}$~, ~$1\leq j < i \leq n$. These variables generate a
flag manifold of the matrix  quantum group  ~$\an$~ which is
dual to ~$\un$~.  ~For generic $r_i\in\bbc$ the representations
$\hat\pi_{\br}$ are irreducible. We give the values of $r_i$ 
when the representations $\hat\pi_{\br}$ are  reducible. 
It is in the latter cases that there arise various
partial equivalences among these representations. 
These partial equivalences are
realized by ~$q$ - difference intertwining operators for which we
give a canonical derivation following \Dob. For $q=1$ these operators 
become the invariant differential operators mentioned above. 
We should also note  that our considerations below are for
general $n\geq 2$, though the case $n=2$, while being done first
as a toy model \ref\Dod{V.K. Dobrev, unpublished, (May 1993).},
is not interesting from the non-commutative point of view since
it involves functions of one variable, and furthermore the
representations and the only possible $q$-difference intertwining
operator are known for $U_q(sl(2))$, (though derived by a
different method), \ref\GP{A.Ch. Ganchev and V.B. Petkova, Phys.
Lett {\bf 233B} (1989) 374-382.}. 

The paper is organized as follows. In Section 2 we recall the
matrix quantum group $GL_q(n)$ and its dual quantum algebra 
$\cu_g$. In Section 3 we give the explicit construction of
representations of $\cu_g$ and its semisimple part $\un$. In
Section 4 we give the reducibility conditions for these
representations and the procedure for the construction of the 
$q$ - difference intertwining operators. In Section 5 we consider
in more detail the case $n=3$.

\vskip 1cm 

\noindent {\bf 2. ~~The matrix   quantum group} 
 
Let us consider an ~$n\times n$~ quantum matrix ~$M$~ with
non-commuting matrix elements ~$a_{ij}$, $1 \leq i,j \leq n$. 
The matrix quantum group ~$\ag ~=~ GL_q(n)$, ~$q\in\bbc$, ~is generated
by the matrix elements ~$a_{ij}$~ with the following
commutation relations \ref\Ma{Yu.I. Manin, Quantum groups 
and non-commutative
geometry, Montreal University preprint, CRM-1561 (1988); ~
Comm. Math. Phys. {\bf 123} (1989) 163-175.}~ 
($\l = q-q^{-1}$)~: \eqna\1 
$$\eqalignno{ a_{ij} a_{i\ell} ~&=~ q^{-1} a_{i\ell} a_{ij} ~, ~{\rm
for} ~ j < \ell ~, &\1 a\cr  
a_{ij} a_{kj} ~&=~ q^{-1} a_{kj} a_{ij} ~, ~{\rm for} ~ i < k ~, &\1
b \cr 
 a_{i\ell} a_{kj} ~&=~ a_{kj} a_{i\ell} ~, ~{\rm for} ~ i < k
~, ~j < \ell ~, &\1 c \cr  
 a_{k\ell} a_{ij} &-  a_{ij} a_{k\ell} ~=~ \l
a_{i\ell} a_{kj} ~, ~{\rm for} ~ i < k ~, ~j < \ell ~.
&\1 d \cr }$$ 
Considered as a bialgebra, it has the following
comultiplication ~$\d_\ca$~ and 
counit ~$\ve_\ca$~:  
\eqn\2{\d_\ca(a_{ij}) ~=~ \sum_{k=1}^n a_{ik} \otimes a_{kj} ~, 
~~~\ve_\ca(a_{ij}) ~=~ \d_{ij} ~. }
This algebra has determinant ~$\cd$~ given by \Ma: 
\eqn\det{ \cd ~=~ \sum_{\r \in S_n} ~
\eps (\r) ~ a_{1,\r (1)} \ldots a_{n,\r (n)}  ~=~ 
\sum_{\r \in S_n} ~\eps (\r) ~ a_{\r (1),1} \ldots a_{\r (n),n} ~
 ~, } 
where summations are over all permutations $\r$ of $\{1, \ldots,
n\}$ and the quantum signature is: 
\eqn\qs{ \eps (\r) ~=~ \prod_{j<k \atop \r (j) >
\r (k) } ~(- q^{-1} )  ~. } 
The determinant obeys \Ma: 
\eqn\deta{ \d_\ca(\cd) ~=~ \cd \otimes \cd ~, ~~~\ve_\ca (\cd)
~=~ 1 ~. } 
The determinant is  central, i.e., it commutes 
with the elements $a_{ik}$~ \Ma: 
\eqn\detc{ a_{ik} ~\cd ~=~ \cd ~a_{ik} ~. } 
Further, if ~$\cd \neq 0$~ one extends the algebra by an element
~$\cd^{-1}$~ which obeys \Ma:  
\eqn\deti{\cd \cd^{-1} ~=~ \cd^{-1} \cd ~=~ 1_\ca ~. } 
 
Next one defines the left and right quantum cofactor matrix 
~$A_{ij}$~ \Ma:  
\eqn\cof{ \eqalign{ 
A_{ij} ~=&~ \sum_{\r (i) = j} ~ { \eps (\r \circ
\s_i ) \over \eps ( \s_i) } ~ a_{1,\r (1)} \ldots {\widehat
a_{ij}} \ldots a_{n,\r (n)} ~= \cr =&
~ \sum_{\r (j) = i} ~ { \eps (\r \circ
\s'_j ) \over \eps ( \s'_j) } ~ a_{\r (1),1} \ldots {\widehat
a_{ij} } \ldots a_{\r (n),n} ~, \cr } }
where $\s_i$ and $\s'_j$ denote the cyclic permutations: 
\eqn\perm{ \s_i ~=~ \{ i,\ldots,1\} ~, ~~~\s'_j ~=~ \{
j,\ldots,n\} ~, } 
and the notation $ {\hat x }$ indicates that $x$ is to be omited.
Now one can show that \Ma: 
\eqn\uni{ \sum_j ~a_{ij} ~A_{\ell j} ~~=~~ \sum_j ~A_{ji} ~a_{j\ell} ~~=~~
\d_{i\ell} ~\cd ~, } 
and  obtain the left and right inverse \Ma: 
\eqn\inv{ M^{-1} ~~=~~ \cd^{-1} ~ A ~~=~~ A ~ \cd^{-1} ~.} 
Thus, one can introduce the antipode in ~$GL_q(n)$~
\Ma\ : 
\eqn\ant{ \g_\ca (a_{ij}) ~~=~~ \cd^{-1} ~ A_{ji} ~~=~~ A_{ji} ~
\cd^{-1} ~.}  

Next we introduce a basis of $GL_q(n)$ which 
consists of monomials 
\eqn\mon{\eqalign{ 
f ~~=&~~ (a_{21})^{p_{21}} \ldots (a_{n,n-1})^{p_{n,n-1}} 
(a_{11})^{\ell_{1}} \ldots (a_{nn})^{\ell_{n}}
(a_{n-1,n})^{n_{n-1,n}} \ldots (a_{12})^{n_{12}} ~= \cr 
=&~~ f_{\bar \ell , \bar p, \bar n} ~, \cr} }
where $\bar \ell , \bar p, \bar n$ denote the sets $\{   
\ell_{i}\}$, $\{p_{ij}\}$, $\{n_{ij}\}$, resp., 
$\ell_{i}, p_{ij}, n_{ij} \in \bbz_+$ and we have used the
so-called normal ordering of the elements $a_{ij}$. Namely, we
first put the elements $a_{ij}$ with $i>j$
in lexicographic order, i.e., if $i<k$ then $a_{ij}$ ($i>j$)
is before $a_{k\ell}$ ($k>\ell$) and $a_{ti}$ ($t>i$) is before
$a_{tk}$ ($t>k$); ~then we put the elements $a_{ii}$;  
~finally we put the elements $a_{ij}$ with
$i<j$ in antilexicographic order, i.e., if $i>k$ then $a_{ij}$
($i<j$) is before $a_{k\ell}$ ($k<\ell$) and $a_{ti}$ ($t<i$) is
before $a_{tk}$ ($t<k$).  
Note that the basis \mon\ icludes also the unit element $1_\ag$
of $\ag$ when all $\{\ell_{i}\}$, $\{p_{ij}\}$, $\{n_{ij}\}$ 
are equal to zero, i.e.:
\eqn\monu{ f_{ \bar 0 , \bar 0, \bar 0} ~~=~~ 1_\ag ~. }

We need the dual algebra of $GL_q(n)$. This is the algebra 
~$\cu_g = U_q(sl(n)) \otimes U_q(\cz)$, where $U_q(\cz)$ is 
central in $\cu_g$ \ref\DP{
V.K. Dobrev and P. Parashar, J. Phys. A: Math. Gen. 
{\bf 26} (1993) 6991.}.  
Let us denote the Chevalley generators of $sl(n)$ by $H_i$, 
$X^\pm_i$, $i = 1, \ldots, n-1$. Then we take for 
the 'Chevalley' generators 
of ~$\cu = U_q(sl(n))$~:~ $k_i = q^{H_i/2}$, $k_i^{-1} = q^{-H_i/2}$, 
$X^\pm_i$, $i = 1, \ldots, n-1$, 
with the following ~{\it algebra}~ relations:      
\eqna\wb 
$$\eqalignno{ &k_i k_j ~=~ k_j k_i ~, \quad 
k_i k^{-1}_i ~=~ k^{-1}_i k_i ~=~ 1_{\cu_g} ~, \quad 
k_i X^\pm_j  ~=~ q^{\pm c_{ij}} X^\pm_j k_i &\wb a\cr  
&[X^+_i , X^-_j ] ~=~ \d_{ij} \left( k^2_i - k^{-2}_i \right)/\l  
~, &\wb b\cr 
&\left(X^\pm_i\right)^2 X^\pm_j ~-~ [2]_q 
 X^\pm_i X^\pm_j X^\pm_i ~+~ X^\pm_j  \left(X^\pm_i\right)^2 
~~=~~ 0 ~, ~~~~\vr i- j\vr =1 ~, &\wb c\cr 
&[X^\pm_i , X^\pm_j ] ~=~ 0 ~, ~~\vr i- j\vr \neq 1 ~, &\wb d\cr 
} $$ 
where ~$c_{ij}$~ is the Cartan matrix of ~$sl(n)$, 
~and ~{\it coalgebra}~ relations: \eqna\wg 
$$\eqalignno{ \d_\cu (k^\pm_i) ~&=~ k^\pm_i \otimes k^\pm_i ~, &\wg a \cr 
\d_\cu (X^\pm_i) ~&=~ X^\pm_i \otimes k_i ~+~ 
k^{-1}_i \otimes X^\pm_i ~, &\wg b\cr 
\ve_\cu (k^\pm_i) ~&=~ 1 ~, \quad \ve_\cu (X^\pm_i) ~=~ 0 
~, &\wg c\cr  
\g_\cu (k_i) ~&=~ k_i^{-1} ~, 
\quad \g_\cu (X^\pm_i) ~=~ - q^{\pm 1} ~X^\pm_i ~, &\wg d\cr}$$
where $k^+_i = k_i$, $k^-_i = k^{-1}_i$.   
Further, we denote the generator of $\cz$ by $H$ and the 
generators of $U_q(\cz)$ by $k = q^{H/2}$, $k^{-1} = q^{-H/2}$,
$k k^{-1} ~=~ k^{-1} k ~=~ 1_{\cu_g}$. 
The generators $k, k^{-1}$ 
commute with the generators of $\cu$, and their coalgebra relations 
are as those of any $k_i$. From now on we shall give most formulae 
only for the generators $k_i, ~X^\pm_i, ~k$, since the analogous formulae 
for $k_i^{-1}, ~k^{-1}$ follow trivially from those for 
$k_i, ~k$, resp.

The bilinear form giving the duality between ~$\cu_g$~ and ~$\ag$~ 
is given by \DP: 
\eqna\pai  
$$\eqalignno{\lg ~k_i ~, ~a_{j\ell} ~\rg ~~&=  ~~ \d_{j\ell} ~q^{(\d_{ij} - 
\d_{i,j+1})/2} ~, &\pai a\cr 
\lg ~X^+_i ~, ~a_{j\ell} ~\rg ~~&=  ~~ \d_{j+1,\ell} \d_{ij} ~, &\pai b\cr 
\lg ~X^-_i ~, ~a_{j\ell} ~\rg ~~&=  ~~ \d_{j-1,\ell} \d_{i\ell} ~, &\pai c\cr 
\lg ~k ~, ~a_{j\ell} ~\rg ~~&=  ~~ \d_{j\ell} ~q^{1/2} ~. &\pai d\cr}$$ 
The pairing between arbitrary elements of ~$\cu_g$~ and 
$f$ follows then from the 
properties of the duality pairing. All this is given in \DP\ and is not 
reproduced here since we shall not need these formulae. 
The pairing \pai{} is standardly supplemented   with 
\eqn\du{ \lg ~y ~, ~1_{\ag} ~\rg ~=~ \ve_{\cu_g} (y)  ~. }

It is well know that the pairing provides the fundamental
representation  of ~$\cu_g$~:
\eqn\fun{ F (y)_{j\ell} ~~=~~ \lg ~y ~, ~a_{j\ell} ~\rg ~, \quad 
y ~=~ k_i, X^\pm_i , k ~. }
Of course, ~$F(k) = q^{1/2} I_n$~, ~where ~$I_n$ ~is the unit
~$n\times n$  ~matrix. 

\vskip 1cm 

\noindent {\bf 3. ~~Representations of ~$\cu_g$~ and ~$\cu$ }

We begin by defining ~{\it two actions}~ of the dual algebra
~$\cu_g$~ on the basis \mon\ of ~$\ag$. 

First we introduce the ~{\it left regular representation}~ 
of ~$\cu_g$~ which
in the $q=1$ case is the infinitesimal version of~: 
\eqn\lrcl{ \pi (Y) \ M ~~=~~ Y^{-1}\ M ~, \quad Y,M ~\in ~GL(n)  ~. } 
Explicitly, we define the action of ~$\cu_g$~ 
as follows (cf. \fun): 
\eqn\lrm{ \pi (y)~ a_{i\ell} ~\doteq ~ 
\left( F \left(y^{-1}\right) M \right)_{i\ell} 
~=~ \sum_j ~F \left(y^{-1}\right)_{ij} ~a_{j\ell}  
~=~ \sum_j ~\lg ~y^{-1}~ , ~a_{ij}~ \rg  ~a_{j\ell} ~, }
where ~$y$~ denotes the generators of ~$\cu_g$~ and ~$y^{-1}$~ is 
symbolic notation, the possible pairs being given explicitly by:
\eqn\lrma{ (y, y^{-1}) ~~~=~~~ (k_i , k_i^{-1}), ~~ 
(X^\pm_i , ~-X^\pm_i), ~~(k, k^{-1}) ~. }
From \lrm\ we find the explicit action of 
the  generators of ~$\cu_g$~: \eqna\lr   
$$\eqalignno{  \pi (k_i)~ a_{j\ell} ~&=~ q^{(\d_{i+1,j} - 
\d_{ij})/2} ~a_{j\ell} ~, &\lr a\cr 
  \pi (X^+_i)~ a_{j\ell} ~&=~ - \d_{ij} ~a_{j+1\ell} ~, 
&\lr b\cr 
  \pi (X^-_i)~ a_{j\ell} ~&=~ - \d_{i+1,j} ~a_{j-1\ell} ~, 
&\lr c\cr 
  \pi (k)~ a_{j\ell} ~&=~ q^{-1/2} ~a_{j\ell} ~. &\lr d\cr}$$ 
The above is supplemented with the following action on the unit
element of $\ag$:
\eqn\lru{ 
\pi (k_i) ~1_\ag ~~=~~ 1_\ag ~, \quad  \pi (X^\pm_i) ~1_\ag ~~=~~
0 ~, \quad  \pi (k) ~1_\ag ~~=~~ 1_\ag ~. } 

In order to derive the action of $\pi(y)$ on arbitrary elements of
the basis \mon, we use the twisted derivation rule consistent with the
coproduct and the representation structure, namely, we take:
~$\pi (y)\vf\psi ~=~ \pi (\d'_{\cu_g}(y))(\vf\otimes \psi)$, 
~where ~$\d'_{\cu_g} ~=
~\sigma \circ \d_{\cu_g}$ ~is the opposite coproduct, 
($\sigma$ is the permutation
operator). Thus, we have: \eqna\tw  
$$\eqalignno{ \pi (k_i)\vf\psi ~~&=~~ \pi (k_i)\vf\cdot \pi
(k_i)\psi\ , &\tw a\cr 
\pi (X^\pm_i)\vf\psi ~~&=~~ \pi (X^\pm_i)\vf\cdot \pi (k^{-1}_i)\psi
~+~ \pi (k_i)\vf\cdot \pi (X^\pm_i)\psi\ , &\tw b\cr 
\pi (k)\vf\psi ~~&=~~ \pi (k)\vf\cdot \pi
(k)\psi\ . &\tw c\cr}$$ 

From now on we suppose that ~$q$~ is not a nontrivial root of unity. 
Applying the above rules one obtains: \eqna\act 
$$\eqalignno{  \pi (k_i)~ (a_{j\ell})^n ~&=~ q^{n(\d_{i+1,j} - 
\d_{ij})/2} ~(a_{j\ell})^n ~, &\act a\cr 
  \pi (X^+_i)~ (a_{j\ell})^n ~&=~ - \d_{ij} ~c_n ~ (a_{j\ell})^{n-1}
 ~a_{j+1\ell} ~, 
&\act b\cr 
  \pi (X^-_i)~ (a_{j\ell})^n ~&=~ - \d_{i+1,j} ~c_n 
~a_{j-1\ell} ~ (a_{j\ell})^{n-1} ~, &\act c\cr 
  \pi (k)~ (a_{j\ell})^n ~&=~ q^{-n/2} ~(a_{j\ell})^n ~, &\act d\cr}$$ 
where 
\eqn\cof{ c_n ~=~ q^{(n-1)/2}
~[n]_q ~, ~~~[n]_q = (q^n - q^{-n})/\l \ . } 
Note that \lru\ and \lr{} are partial cases of \act{} for $n=0$
and $n=1$ resp. (cf. \monu).

\vskip 5mm

Analogously, we introduce the ~{\it right action}~ 
(see also \ref\MNS{T. Masuda, K. Mimachi,
Y. Nakagami, M. Noumi, Y. Sabuti and K. Ueno, Lett. Math. Phys.
{\bf 19} (1990) 187-194; Lett. Math. Phys. {\bf 19} (1990)
195-204.}) which in 
the classical case is the infinitesimal counterpart of~:
\eqn\lrcr{ \pi_R (Y) ~ M ~~=~~ M ~ Y ~, \quad Y,M ~\in ~GL(n)  ~. } 
Thus, we define the right action of ~$\cu_g$~ 
as follows (cf. \fun): 
\eqn\lrr{ \pi_R (y)~ a_{i\ell} ~=~ \left(  M F (y) \right)_{i\ell} 
~=~ \sum_j ~a_{ij} ~F (y)_{j\ell} ~ 
~=~ \sum_j ~a_{ij} ~ \lg ~y~ , ~a_{j\ell}~ \rg ~, }
where ~$y$~ denotes the generators of ~$\cu_g$~. 

From \lrr\ we find the explicit right action of 
the  generators  of ~$\cu_g$~: \eqna\lrrr   
$$\eqalignno{  \pi_R (k_i)~ a_{j\ell} ~&=~ q^{(\d_{i\ell} - 
\d_{i+1,\ell})/2} ~a_{j\ell} ~, &\lrrr a\cr 
  \pi_R (X^+_i)~ a_{j\ell} ~&=~ \d_{i+1,\ell} ~a_{j,\ell -1} ~, 
&\lrrr b\cr 
  \pi_R (X^-_i)~ a_{j\ell} ~&=~ \d_{i\ell} ~a_{j,\ell +1} ~, 
&\lrrr c\cr 
  \pi_R (k)~ a_{j\ell} ~&=~ q^{1/2} ~a_{j\ell} ~, &\lrrr d\cr}$$ 
supplemented by the right action on the unit element:
\eqn\lrur{\pi_R (k_i) ~1_\ag ~=~ 1_\ag ~, \quad  \pi_R (X^\pm_i)
~1_\ag ~~=~~ 0 ~, \quad  \pi_R (k) ~1_\ag ~~=~~ 1_\ag ~. }

The twisted derivation rule is now given by ~$\pi_R (y)\vf\psi
~=~ \pi_R (\d_{\cu_g}(y))(\vf\otimes \psi)$, i.e., \eqna\twr 
$$\eqalignno{ \pi_R (k_i)\vf\psi ~&=~ \pi_R (k_i)\vf\cdot \pi_R
(k_i)\psi\ ,&\twr a\cr 
\pi_R (X^\pm_i)\vf\psi ~&=~ \pi_R (X^\pm_i)\vf\cdot \pi_R (k_i)\psi
+\pi_R (k_i^{-1})\vf\cdot \pi_R (X^\pm_i)\psi\ ,&\twr b\cr 
\pi_R (k)\vf\psi ~&=~ \pi_R (k)\vf\cdot \pi_R 
(k)\psi\ ,&\twr c\cr 
}$$
Using this, we find: \eqna\ract  
$$\eqalignno{  \pi_R (k_i)~ (a_{j\ell})^n ~&=~ q^{n(\d_{i\ell} - 
\d_{i+1,\ell})/2} ~(a_{j\ell})^n ~, &\ract a\cr 
  \pi_R (X^+_i)~ (a_{j\ell})^n ~&=~  \d_{i+1,\ell} ~c_n ~a_{j,\ell -1} 
~ (a_{j\ell})^{n-1}   ~, &\ract b\cr 
  \pi_R (X^-_i)~ (a_{j\ell})^n ~&=~ \d_{i\ell} ~c_n 
 ~ (a_{j\ell})^{n-1} ~a_{j,\ell+1} ~, &\ract c\cr 
  \pi_R (k)~ (a_{j\ell})^n ~&=~ q^{n/2} ~(a_{j\ell})^n ~. 
&\ract d\cr}$$

Let us now introduce the elements ~$\vf$~ as formal power series of the 
basis \mon: 
\eqn\ser{\eqalign{
\vf ~=&~ \sum_{{\bar \ell ,\bar m,\bar n \in\bbz_+}}
\mu_{{\bar \ell , \bar m, \bar n }} ~
(a_{21})^{m_{21}} \ldots (a_{n,n-1})^{m_{n,n-1}} 
(a_{11})^{\ell_{1}} \ldots (a_{nn})^{\ell_{n}} \times \cr &\times 
(a_{n-1,n})^{n_{n-1,n}} \ldots (a_{12})^{n_{12}} ~. \cr}}
 
By \act{} and \ract{} we have defined left and right action of ~$\cu_g$~
on ~$\vf$. 
As in the classical case the left and right actions 
commute, and as in \Dob\ we shall use the right action to reduce
the left regular representation (which is highly reducible). In
particular, we would like the right 
action  to mimic some properties of a highest weight module, 
i.e., annihilation by the raising generators $X^+_i$ and scalar
action by the (exponents of the) Cartan operators $k_i,k$. In the
classical case these properties are also called right covariance
\Dob. However, first we have to make a change of basis using the 
$q$-analogue of the classical Gauss decomposition. 
For this we have to 
suppose that the principal minor determinants of $M$~:  
\eqn\prm{ 
\eqalign{ \cd_m ~&=~ \sum_{\r \in S_m} ~
\eps (\r) ~ a_{1,\r (1)} \ldots a_{m,\r (m)}  ~=\cr &=~  
\sum_{\r \in S_m} ~\eps (\r) ~ a_{\r (1),1} \ldots a_{\r (m),m} ~
~, \quad m\leq n ~, \cr } } 
are invertible; note that ~$\cd_n = \cd$, ~$\cd_{n-1} = A_{nn}$. 
Thus, using \uni\ for $i=\ell=n$ we can express, e.g.,  
$a_{nn}$ in terms of other elements: 
\eqn\ann{ a_{nn} ~=~ \left( \cd ~-~ \sum_{j<n} ~a_{nj} ~A_{nj} \right) 
~ \cd_{n-1}^{-1} ~=~ \cd_{n-1}^{-1} ~\left( \cd ~-~ 
\sum_{j<n} ~A_{jn} ~a_{jn} \right)  
~. }

Further, for the ordered sets 
$I=\{ i_1 < \cdots < i_r \}$ and $J=\{ j_1 < \cdots < j_r \}$,
let  $\xi^I_J$ be the  $r$-minor determinant
with respect to rows $I$ and columns $J$ such that 
\eqn\mnr{\xi^I_J ~=~ \sum_{\r\in S_r} \eps(\r)~
a_{i_{\r(1)}j_1}\cdots a_{i_{\r(r)}j_r} ~.} 
Note that $\xiL{i  }{i  } = \cd_i$~. ~
Then one has \ref\ANO{H. Awata, M. Noumi and S. Odake, 
preprint YITP/K-1016 (1993).} ($i,j,\ell = 1,\ldots,n$)~:  
\eqn\gau{ 
a_{i\ell} ~=~ \sum_j B_{ij}Z_{j\ell} ~,\quad
B_{i\ell} ~=~ \xiL{\ell-1\,i}{\ell     } \cd_{\ell-1}^{-1} ~,\quad
Z_{i\ell} ~=~ \cd_i^{-1} \xiL{i     }{i-1\,\ell} ~,}
$B_{i\ell}=0$ for $i<\ell$, ~$Z_{i\ell}=0$ for $i>\ell$, 
(which follows from the obvious extension of \mnr\ to the case
when $I$, resp. $J$, is not ordered). Then ~$Z_{ij}$, $i<j$, ~may
be regarded as a $q$-analogue of local coordinates of the flag
manifold ~$B\backslash GL(n)$. 

For our purposes we need a refinement of this decomposition: 
\eqn\gua{B_{i\ell} ~~=~~ Y_{i\ell} D_{\ell\ell} ~, \quad 
Y_{i\ell} ~=~ \xiL{\ell-1\,i}{\ell     } \cd_{\ell}^{-1} ~,\quad 
D_{\ell\ell} ~=~ \cd_{\ell} \cd_{\ell -1}^{-1} ~, 
\quad (\cd_0 \equiv 1_\ag) ~,} 
where ~$Y_{j\ell}$, $j>\ell$, ~may be regarded as a $q$-analogue of local
coordinates of the flag manifold ~$GL(n)/DZ$.

Clearly, we can replace the basis \mon\ of $\ag$ with a basis in 
terms of $Y_{i\ell}$, $i>\ell$, $\cd_{\ell}$, $Z_{i\ell}$, 
$i<\ell$. (Note that $Y_{ii} = Z_{ii} = 1_\ag$.) We could have used 
also $D_{\ell\ell}$ instead of $\cd_\ell$, but this choice is more 
convenient since below we shall impose $\cd_n = \cd = 1_\ag$. Thus, we 
consider formal power series: 
\eqn\sera{\eqalign{
\vf ~=&~ \sum_{{ \bar m,\bar n \in\bbz_+
\atop \bar \ell \in\bbz}}
\mu'_{{\bar \ell , \bar m, \bar n }} ~
(Y_{21})^{m_{21}} \ldots (Y_{n,n-1})^{m_{n,n-1}} 
(\cd_{1})^{\ell_{1}} \ldots (\cd_{n})^{\ell_{n}} \times \cr &\times 
(Z_{n-1,n})^{n_{n-1,n}} \ldots (Z_{12})^{n_{12}} ~. \cr}}

Now, let us impose right covariance \Dob\ with respect to ~$X^+_i$~, 
i.e., we require: 
\eqn\rcf{ \pi_R (X^+_i) ~\vf ~~=~~ 0 ~. } 
First we notice that:
\eqn\rcfa{ \pi_R (X^+_i) ~ \xi^I_J ~~=~~ 0 ~, \quad {\rm for}~~ 
J ~=~ \{1,\ldots,j\} ~, ~\forall ~I ~, }  
from which follow:  
\eqn\rcff{ \pi_R (X^+_i) ~\cd_j ~~=~~ 0 ~,  
\quad \pi_R (X^+_i) ~Y_{j\ell} ~~=~~ 0 ~.}
On the other hand ~$\pi_R (X^+_i)$~ acts nontrivially on ~$Z_{j\ell}$~. ~
Thus, \rcf\ simply means that our functions ~$\vf$~ do not  depend
on ~$Z_{j\ell}$~. ~Thus, the functions obeying \rcf\ are: 
\eqn\seras{
\vf ~=~ \sum_{{\bar \ell \in\bbz ~, ~ \bar m \in\bbz_+  }}
\mu_{{\bar \ell , \bar m}} ~
(Y_{21})^{m_{21}} \ldots (Y_{n,n-1})^{m_{n,n-1}} 
(\cd_{1})^{\ell_{1}} \ldots (\cd_{n})^{\ell_{n}} 
~. }

Next, we impose right covariance with respect to ~$k_i,k$~: 
\eqna\rck
$$\eqalignno{
&\pi_R (k_i) ~\vf ~~=~~ q^{r_i/2} ~\vf ~, &\rck a\cr
&\pi_R (k) ~\vf ~~=~~ q^{\hr/2} ~\vf ~, &\rck b\cr}$$
where $r_i,\hr$ are parameters to be specified below. 
On the other hand using \twr{a,c}, \ract{a,c} we  have: 
\eqn\rcfka{\pi_R (k_i) ~\xi^I_J ~~=~~ q^{\d_{ij}/2}~\xi^I_J ~, 
\quad \pi_R (k) ~\xi^I_J ~~=~~ q^{j/2}~\xi^I_J ~, 
\quad {\rm for}~~ J ~=~ \{1,\ldots,j\} ~, ~\forall ~I ~, }    
from which follows: 
\eqna\rcfk
$$\eqalignno{
&\pi_R (k_i) ~\cd_j ~~=~~ q^{\d_{ij}/2}~ \cd_j ~, \quad 
\pi_R (k) ~\cd_j ~~=~~ q^{j/2}~ \cd_j ~, &\rcfk a\cr
&\pi_R (k_i) ~Y_{j\ell} ~~=~~ Y_{j\ell} ~, \quad
\pi_R (k) ~Y_{j\ell} ~~=~~ Y_{j\ell} ~, &\rcfk b\cr}$$
and thus we have: 
\eqna\rk
$$\eqalignno{&\pi_R(k_i) ~\vf ~=~ q^{\ell_i/2}~ \vf  ~, &\rk a\cr  
&\pi_R(k) ~\vf ~=~ q^{\sum_{j=1}^n j\ell_j/2}~ \vf  ~. &\rk b\cr}$$  

Comparing right covariance conditions \rck{} with the 
direct calculations \rk{} we obtain 
$\ell_i = r_i$, for $i<n$, $\sum_{j=1}^n j\ell_j = \hr$. This 
means that  ~$r_i,\hr\in\bbz$~ and that there is no summation 
in $\ell_i$, also $\ell_n = (\hr - \sum_{i=1}^{n-1} ir_i)/n$. 
 
Thus, the reduced functions obeying \rcf\ and \rck\ are:
\eqn\seras{
\vf ~=~ \sum_{{ \bar m \in\bbz_+}} 
\mu_{\bar m} ~
(Y_{21})^{m_{21}} \ldots (Y_{n,n-1})^{m_{n,n-1}} 
(\cd_{1})^{r_{1}} \ldots (\cd_{n-1})^{r_{n-1}} 
(\cd_{n})^{{\hat \ell}} ~, }
where $  {\hat \ell} ~=~ (\hr - \sum_{i=1}^{n-1} ir_i)/n $. 

Next we would like to derive the ~$\cu_g$~ - action ~$\pi$~ on
~$\vf$~.  ~First, we notice that ~$\cu$~  acts trivially on
$\cd_n = \cd$~: 
\eqn\actd{ \pi (X^\pm_i) ~\cd ~~=~~ 0 ~, \quad 
\pi(k_i) ~ \cd ~~=~~ \cd ~.} 
Then we note: 
\eqn\acad{ \pi (k) ~\cd_j ~~=~~ q^{-j/2} ~\cd_j ~, \quad 
\pi (k) ~Y_{j\ell} ~~=~~ Y_{j\ell} ~, } 
from which follows: 
\eqn\acbd{ \pi (k) ~\vf ~~=~~ q^{-\hr/2} ~\vf ~.} 
Thus, the action of ~$\cu$~ involves only the parameters 
~$r_i$, $i<n$, while the action of ~$U_q(\cz)$~ involves 
only the parameter ~$\hr$. 
Thus we can consistently also from the representation 
theory point of view restrict to the  
matrix quantum group $SL_q(n)$, i.e., we set: 
\eqn\sl{\cd ~ =~ \cd^{-1} ~=~ 1_\ag ~. }  
Then the dual algebra is $\cu = U_q(sl(n))$. 
This is justified as in the $q=1$ case \Dob\ since for our
considerations only the semisimple part of the algebra is
important. (This would not be possible for  
the multiparameter deformation of $GL(n)$ 
\ref\Sua{A. Sudbery, J. Phys. A : Math. Gen. {\bf 23} (1990) L697.}, 
\ref\Schi{A. Schirrmacher, Zeit. f. Physik {\bf C50} (1991) 321.}, 
since there $\cd$ is not central. 
Nevertheless, we expect most of the essential features 
of our approach to be preserved 
since the dual algebra can be transformed 
as a commutation algebra to the one-parameter 
$\cu_g$, with the extra parameters entering only the co-algebra 
structure \DP.)

Thus, the reduced functions for the ~$\cu$~ action are: 
\eqna\serasa
$$\eqalignno{
\tv ({\bar Y}, {\bar \cd}) 
~=&~ \sum_{{ \bar m \in\bbz_+}} 
\mu_{\bar m} ~
(Y_{21})^{m_{21}} \ldots (Y_{n,n-1})^{m_{n,n-1}} 
(\cd_{1})^{r_{1}} \ldots (\cd_{n-1})^{r_{n-1}} ~= &\serasa a\cr 
~=& ~~\hp ({\bar Y})~ 
(\cd_{1})^{r_{1}} \ldots (\cd_{n-1})^{r_{n-1}} ~, &\serasa b\cr}$$   
where ${\bar Y}, {\bar \cd}$ denote the variables $Y_{il}$, $i>\ell$, 
$\cd_i$, $i<n$. Next  we  calculate:
\eqna\accd \eqna\acdd 
$$\eqalignno{ &\pi (k_i) ~\cd_j ~~=~~ q^{-\d_{ij}/2} ~\cd_j ~, 
&\accd a\cr 
&\pi (X^+_i) ~\cd_j ~~=~~ -\d_{ij} ~Y_{j+1,j} ~\cd_j ~, 
&\accd b\cr 
&\pi (X^-_i) ~\cd_j ~~=~~ 0 ~, &\accd c \cr 
&\pi (k_i) ~Y_{j\ell} ~~=~~ 
q^{{1\over 2}(\d_{i+1,j} - \d_{ij} - \d_{i+1,\ell} + \d_{i\ell})} 
~Y_{j\ell} &\acdd a\cr 
&\eqalign{\pi (X^+_i) ~Y_{j\ell} ~~=&~~ -\d_{ij} ~Y_{j+1,\ell} ~+~ 
\d_{i\ell} ~ q^{1 - \d_{j,\ell +1}/2}  
~Y_{\ell +1, \ell} ~Y_{j\ell} ~+\cr ~&+~ ~  
\d_{i+1,\ell} ~ \left( q^{-1} ~Y_{j,\ell -1} ~-~ Y_{\ell,\ell -1} ~
Y_{j\ell} \right)  ~, \cr} &\acdd b\cr 
&\pi (X^-_i) ~Y_{j\ell} ~~=~~ - \d_{i+1,j} ~ q^{-\d_{i\ell}/2} ~
Y_{j-1,\ell}   ~. &\acdd c \cr } $$

These results have the important consequence that the 
degrees of the variables ~$\cd_j$~ are not changed by the 
action of ~$\cu$. Thus, the parameters ~$r_i$~ indeed characterize 
the action of ~$\cu$~, ~i.e., we have obtained representations 
of ~$\cu$.  We shall denote 
by ~$\cc_{\br}$~ the representation space of functions in 
\serasa{} which have covariance properties \rcf, \rck{a}, 
and the representation acting in ~$\cc_{\br}$~ we denote  
by ~$\tilde\pi_{\br}$ - here  a renormalization
of the explicit formulae may be done to simplify things. 
To obtain this  representation more explicitly one just 
applies \accd{}, \acdd{} to the basis in \serasa{} using 
\tw{}. In particular, we have: \eqna\abcd
$$\eqalignno{ \pi (k_i) ~(\cd_j)^n ~~=&~~ q^{-n\d_{ij}/2} ~(\cd_j)^n ~, 
\quad n\in\bbz ~, &\abcd a\cr 
\pi (X^+_i) ~(\cd_j)^n ~~=&~~ -\d_{ij} ~\bc_n  ~ 
~Y_{j+1,j} ~(\cd_j)^n ~, \quad n\in\bbz ~, &\abcd b\cr 
\pi (X^-_i) ~(\cd_j)^n ~~=&~~ 0 ~, \quad n\in\bbz ~, 
&\abcd c \cr 
\eqna\abdd
\pi (k_i) ~(Y_{j\ell})^n ~~=&~~ 
q^{{n\over 2}(\d_{i+1,j} - \d_{ij} - \d_{i+1,\ell} + \d_{i\ell})} 
~(Y_{j\ell})^n ~, \quad n\in\bbz_+ ~, 
&\abdd a\cr 
\pi (X^+_i) ~(Y_{j\ell})^n ~~=&~~ 
-\d_{ij} ~c_n ~(Y_{j\ell})^{n-1} ~Y_{j+1,\ell} ~+&\cr &~~+~   
\d_{i\ell} ~ q^{1  -n\d_{j,\ell +1}/2} ~c_n   
~Y_{\ell +1, \ell} ~(Y_{j\ell})^n ~+&\cr &~~+ ~  
\d_{i+1,\ell} ~ \bc_n ~\left( q^{-1} ~Y_{j,\ell -1} ~(Y_{j\ell})^{n-1}
~-~ Y_{\ell,\ell -1} ~ Y_{j\ell}^n \right), \quad n\in\bbz_+ 
&\abdd b\cr 
\pi (X^-_i) ~(Y_{j\ell})^n ~~=&~~ - \d_{i+1,j} ~ 
q^{-\d_{j,\ell +1}n/2} ~c_n ~
Y_{j-1,\ell} ~(Y_{j\ell})^{n-1}  ~, \quad n\in\bbz_+ ~, 
&\abdd c \cr } $$
where 
\eqn\coe{\bc_n ~~=~~ q^{(1-n)/2}~ [n]_q ~. }

Further, since the action of ~$\cu$~ is not affecting the degrees
of $\cd_i$,  we introduce (as in \Dob)  the restricted functions
~$\hp ({\bar Y})$~ by the
formula which is prompted in \serasa{b}~: 
\eqn\res{ \hp({\bar Y}) ~\equiv~ \bigl( \ca\tv ) ({\bar Y}) ~\doteq ~\tv
({\bar Y}, \cd_1 = \cdots = \cd_{n-1} = 1_\ag) ~. } 
We denote the representation space of ~$\hp({\bar Y})$~ by
~$\tc_{\br}$~ and the representation acting in ~$\tc_{\br}$~
by ~$\hat\pi_{\br}$~. ~Thus, the operator ~$\ca$~ acts from
~$\cc_{\br}$~ to ~$\tc_{\br}$~. ~The properties of
~$\tc_{\br}$~ follow from the intertwining requirement for ~$\ca$~
\Dob:  
\eqn\int{ \hat\pi_{\br} ~\ca ~~=~~ \ca ~ \tilde
\pi_{\br} ~.} 

\vskip 1cm

\noindent {\bf 4. ~~Reducibilty and ~$q$ - difference
intertwining operators}  

We have defined the representations ~$\hat\pi_{\br}$~ for
~$r_i\in\bbz$. However, notice that we 
can consider the restricted functions $\hp({\bar Y})$ for 
arbitrary complex $r_i$. We shall make these extension from now on, 
since this gives the same set of representations for ~$\un$~ 
as in the case $q=1$. 

Now we make some statements which are true in the classical 
case  \Dob, and will be illustrated below. 
For any ~$i,j$, such that  
~$1\leq i \leq j \leq n-1$, define: 
\eqn\rij{m_{ij} ~\equiv ~ r_i + \cdots + r_j + j-i+1 ~, }
note $m_i = m_{ii} = r_i +1$, $m_{ij} ~=~ m_i + \cdots + m_j$. 
Note that the possible choices of $i,j$ are in 1-to-1 
correspondence with the positive roots ~$\a = \a_{ij} = 
\a_i + \cdots + \a_j$ of the root system of $sl(n)$, the cases 
$i=j = 1 \ldots, n-1$ enumerating the simple roots $\a_i = \a_{ii}$. 
In general, ~$m_{ij}\in\bbc$~ for the representations ~$\hat\pi_{\br}$, 
while ~$m_{ij}\in\bbz$~ for the representations ~$\pi_{\br}$.  
If ~$m_{ij} \notin \bbn$~ for all possible $i,j$ 
the representations ~$\hat\pi_{\br}$, $\pi_{\br}$ ~are irreducible. 
If ~$m_{ij} \in \bbn$~ for some $i,j$ the representations 
$\hat\pi_{\br}$, $\pi_{\br}$ ~are reducible. The corresponding 
irreducible subrepresentations are still infinite-dimensional 
unless ~$m_i\in\bbn$~ for all $i=1,\ldots,n-1$. 
The representation spaces of the irreducible subrepresentations 
are invariant irreducible subspaces of our representation spaces. 
These invariant subspaces  are spanned by 
functions depending on all variables ~$Y_{j\ell}$~, 
~except when for some ~$s\in\bbn$, 
$1\leq s \leq n-1$, we have ~$m_s = m_{s+1} = \cdots = m_{n-1} 
=1$. In the latter case these functions depend only on the  
~$(s-1)(2n-s)/2$~ variables ~$Y_{j\ell}$~ with ~$\ell<s$, ~
(the unrestricted subrepresentation functions depend still on $D_\ell$ 
with $\ell<s$). ~In particular, for ~$s=2$~ the
restricted subrepresentation  functions depend only 
on the $n-1$ variables ~$Y_{j1}$. The latter situation is relatively 
simple also in the $q$ case since these variables are $q$-commuting~: 
~$Y_{j1}Y_{k1} ~=~ qY_{k1}Y_{j1}$~, ~$j>k$. 
(For $s=1$ the irreducible subrepresentation is one dimensional, 
hence no dependence on any variables.)

Furthermore, for ~$m_{ij} \in \bbn$~ the 
representation $\hat\pi_{\br}$, $\pi_{\br}$, resp.,  
is partially equivalent to the representation 
$\hat\pi_{\br'}$, $\pi_{\br'}$, resp.,  
with $m'_\ell = r'_\ell +1$ being explicitly given as 
follows \Dob: 
\eqn\redd{ 
m'_\ell ~~=~~ \cases{ 
m_\ell ~, &\quad for~ $\ell \neq i-1,i,j,j+1$ ~, \cr 
m_{\ell j}  ~, & \quad for~ $\ell =i-1$ ~, \cr 
-m_{\ell +1,j} ~, & \quad for~ $\ell =i< j$ ~, \cr 
-m_{i,\ell -1} ~, & \quad for~ $\ell =j> i$ ~, \cr 
-m_\ell   ~, & \quad for~ $\ell =i=j$ ~, \cr 
m_{i\ell} ~, & \quad for~ $\ell =j+1$ ~. \cr}}  

These partial equivalences are realized by intertwining 
operators: \eqna\op 
$$\eqalignno{ &\ci_{ij} ~: ~\cc_{\br} ~\longrightarrow
~\cc_{\br'} ~, \quad m_{ij}\in\bbn  ~, 
&\op a\cr 
&I_{ij} ~: ~\tc_{\br} ~\longrightarrow
~\tc_{\br'} ~, \quad m_{ij}\in\bbn 
 ~,  &\op b\cr }$$
i.e., one has: \eqna\inta 
$$\eqalignno{ & \ci_{ij} \circ \pi_{\br} ~=~ \pi_{\br'}
\circ \ci_{ij}  ~, \quad m_{ij}\in\bbn ~, 
&\inta a\cr 
&I_{ij} \circ \hat\pi_{\br} ~=~ \hat\pi_{\br'}
\circ I_{ij}  ~, \quad m_{ij}\in\bbn 
~. &\inta b\cr }$$  
The invariant irreducible subspace of ~$\hat\pi_{\br}$~ 
(resp. $\pi_{\br}$) discussed above 
is the intersection of the kernels of all 
intertwining operators acting from
~$\hat\pi_{\br}$ ~(resp. $\pi_{\br}$). When all $m_i\in\bbn$ the  
invariant subspace is finite-dimensional with dimension 
~$\prod_{ 1\leq i \leq j \leq n - 1 } ~ m_{ij} ~/ ~ 
\prod_{t = 1}^{n-1} t!$~, and all finite-dimensional irreps 
of $U_q(sl(n))$ can be obtained in this way.

We present now a canonical procedure for the derivation 
of these intertwining operators following the $q=1$ procedure of  \Dob. 
By this procedure one should take as
intertwiners (up to nonzero multiplicative constants):
\eqna\intb 
$$\eqalignno{ &\ci^m_{ij} ~~=~~ \cp^m_{ij} \left( \pi_R (X^-_i), 
\ldots, \pi_R (X^-_j) \right),  ~~~~m=m_{ij} \in\bbn ~, &\intb a\cr 
&I^m_{ij} ~~=~~ \cp^m_{ij} \left( \hat\pi_R (X^-_i), 
\ldots, \hat\pi_R (X^-_j) \right), 
~~~~m=m_{ij} \in\bbn  ~, &\intb b\cr  }$$
where ~$\cp^m_{ij}$~ is a homogeneous 
polynomial in each of its ~$(j-i+1)$~ variables of degree ~$m$. 
This polynomial  gives a singular vector ~$v_{ij}$~ 
in a Verma module $V^{\L(\br)}$ 
with highest weight $\L(\br)$ determined by $\br$, (cf. \Dob),
i.e.:  
\eqn\sing{ v_{ij} ~~=~~ 
\cp^m_{ij} \left( X^-_i, \ldots, X^-_j \right) ~\otimes ~v_0 ~, }
where $v_0$ is the highest weight vector of $V^{\L(\br)}$.   
In particular, in the case of the simple 
roots, i.e., when ~$m_i = m_{ii} = r_i +1 \in\bbn$, 
we have 
\eqna\intc 
$$\eqalignno{ &\ci^{m_i}_{i} ~~=~~  \left( \pi_R (X^-_i) 
\right)^{m_i},  ~~~~m_{i} \in\bbn ~, &\intc a\cr 
&I^{m_i}_{i} ~~=~~ \left( \hat\pi_R (X^-_i) 
\right)^{m_i},  ~~~~m_{i} \in\bbn ~. &\intc b\cr }$$ 
For the nonsimple roots one should use the explicit expressions
for the singular vectors of the Verma modules over ~$\un$~ 
given in \ref\Doc{V.K. 
Dobrev, J. Phys. A: Math. Gen. {\bf 25} (1992) 149.}.  
Implementing the above one should be careful since $\hat\pi_R
(X^-_i)$ is not preserving the reduced spaces ~$\cc_{\br}$,
~$\tc_{\br}$, which is of course a prerequisite for \inta{}, 
\intb{}, \intc{}.  

\vskip 1cm 

\noindent {\bf 5. ~~The case of ~$\uq$}

In this Section we consider in more detail the case $n=3$.  We
could have started (following the chronology) also with the case
$n=2$ involving functions of one variable \Dod. However, though
by a different method, this case was obtained in \GP.  It can
also be obtained by restricting the construction for the
(complexification of the) Lorentz quantum algebra of \DDF\ to one
of its $U_q(sl(2))$ subalgebras.

Let us now for $n=3$ denote the coordinates on the flag 
manifold by: ~$\x ~=~
Y_{21}$, ~$\y ~=~ Y_{32}$, ~$\z ~=~ Y_{31}$. We note for future
use the commutation relations between these coordinates: 
\eqn\coo{ \x\y ~=~ q\y\x - \l\z ~, \quad \y\z ~=~ q\z\y ~, \quad
\z\x ~=~ q\x\z ~.} 

The reduced functions for the ~$\cu$~ action are (cf. \serasa{}): 
\eqna\sersa
$$\eqalignno{
\tv ({\bar Y}, {\bar \cd}) 
~~=&~~ \sum_{{ j,n,\ell \in\bbz_+}} \mu_{j,n,\ell}~ 
\x^j ~ \z^n ~ \y^\ell 
(\cd_{1})^{r_{1}} ~ (\cd_{2})^{r_{2}} 
~= &\sersa a\cr 
=& ~~\sum_{{ j,n,\ell \in\bbz_+}} \mu_{j,n,\ell}~ \tv_{jn\ell} ~,  
&\sersa b\cr
\tv_{jn\ell} ~~=& ~~ 
\x^j ~ \z^n ~ \y^\ell ~
(\cd_{1})^{r_{1}} ~ (\cd_{2})^{r_{2}} ~. &\sersa c\cr
}$$   

Now the action of $\uq$ on \sersa{} is given explicitly by:
\eqna\aba 
$$\eqalignno{ 
\pi (k_1) ~ \tv_{jn\ell} ~~=&~~ q^{j + (n-\ell-r_1)/2} 
~\tv_{jn\ell} ~, &\aba a\cr 
\pi (k_2) ~ ~\tv_{jn\ell}
~~=&~~ q^{\ell  + (n-j-r_2)/2}  ~\tv_{jn\ell} ~, &\aba b\cr 
\pi (X^+_1)  ~\tv_{jn\ell} ~~=&~~ 
q^{(1 + n - \ell - r_1)/2} ~[n+j-\ell -r_1]_q   ~\tv_{j+1,n\ell} 
~+&\cr &~~+~ 
q^{j + (n - \ell - 3r_1 -1)/2} ~[\ell]_q ~  ~\tv_{j,n+1,\ell -1} 
~, &\aba c\cr 
\pi (X^+_2)  ~\tv_{jn\ell} 
~~=&~~ 
q^{(1 + n - j - r_2)/2} ~[\ell -r_2]_q  ~\tv_{jn,\ell +1}  
~-&\cr &~~-~ 
q^{-\ell + (j - n + r_2 -1)/2} ~[j]_q ~  ~\tv_{j-1,n+1,\ell} 
~, &\aba d\cr 
\pi (X^-_1)  ~\tv_{jn\ell} 
~~=&~~ 
q^{(\ell - n + r_1 -1)/2} ~[j]_q  ~\tv_{j-1,n\ell}  
~, &\aba e\cr 
\pi (X^-_2)  ~\tv_{jn\ell} 
~~=&~~ 
-~q^{(n - j + r_2 -1)/2} ~[\ell]_q  ~\tv_{jn,\ell -1}  
~-&\cr &~~-~ 
q^{-\ell + (n - j + r_2 -1)/2} ~[n]_q  ~\tv_{j+1,n-1,\ell}  
~. &\aba f\cr 
} $$

It is easy to check that ~$\pi(k_i)$, $\pi(X^\pm_i)$ 
satisfy \wb{}. 
It is also clear that we can remove the inessential phases  by
setting: 
\eqn\red{\tilde\pi_{r_1,r_2} (k_i) ~~=~~ \pi (k_i) ~, \quad 
\tilde\pi_{r_1,r_2} (X^\pm_i) ~~=~~
q^{\pm (r_i-1)/2} ~\pi (X^\pm_i) ~. }  
Then ~$\tilde\pi_{r_1,r_2}$~  also satisfy \wb{}.  

Then we consider the restricted functions (cf. \res): 
\eqna\sersb
$$\eqalignno{ \hp({\bar Y}) 
~~=&~~ \sum_{{ j,n,\ell \in\bbz_+}} \mu_{j,n,\ell}~ 
\x^j ~ \z^n ~ \y^\ell ~= &\sersb a\cr 
=& ~~\sum_{{ j,n,\ell \in\bbz_+}} \mu_{j,n,\ell}~ \hp_{jn\ell} ~,  
&\sersb b\cr
\hp_{jn\ell} ~~=& ~~ 
\x^j ~ \z^n ~ \y^\ell ~
~. &\sersb c\cr
}$$   

As a consequence of the intertwining property \int\ we obtain that 
$\hp_{jn\ell}$ obey the same transformation rules \aba{} as 
$\tv_{jn\ell}$, i.e., (cf. also \red) we have: \eqna\abb 
$$\eqalignno{ 
\hat\pi_{r_1,r_2} (k_1) ~ \hp_{jn\ell} ~~=&~~ q^{j + (n-\ell-r_1)/2} 
~\hp_{jn\ell} ~, &\abb a\cr 
\hat\pi_{r_1,r_2} (k_2) ~ ~\hp_{jn\ell}
~~=&~~ q^{\ell  + (n-j-r_2)/2}  ~\hp_{jn\ell} ~, &\abb b\cr 
\hat\pi_{r_1,r_2} (X^+_1)  ~\hp_{jn\ell} ~~=&~~ 
q^{(n - \ell )/2} ~[n+j-\ell -r_1]_q   ~\hp_{j+1,n\ell} 
~+&\cr &~~+~ 
q^{j - r_1 -1 + (n - \ell )/2} ~[\ell]_q ~  ~\hp_{j,n+1,\ell -1} 
~, &\abb c\cr 
\hat\pi_{r_1,r_2} (X^+_2)  ~\hp_{jn\ell} 
~~=&~~ 
q^{(n - j)/2} ~[\ell -r_2]_q  ~\hp_{jn,\ell +1}  
~-&\cr &~~-~ 
q^{ r_2 - 1 - \ell + (j - n )/2} ~[j]_q ~  ~\hp_{j-1,n+1,\ell} 
~, &\abb d\cr 
\hat\pi_{r_1,r_2} (X^-_1)  ~\hp_{jn\ell} 
~~=&~~ 
q^{(\ell - n)/2} ~[j]_q  ~\hp_{j-1,n\ell}  
~, &\abb e\cr 
\hat\pi_{r_1,r_2} (X^-_2)  ~\hp_{jn\ell} 
~~=&~~ 
-~q^{(n - j)/2} ~[\ell]_q  ~\hp_{jn,\ell -1}  
~-&\cr &~~-~ 
q^{-\ell + (n - j)/2} ~[n]_q  ~\hp_{j+1,n-1,\ell}  
~. &\abb f\cr 
} $$

Let us introduce the following operators acting on our functions: 
\eqna\opes 
$$\eqalignno{ 
\hm^\pm_\k ~\hp({\bar Y}) ~~=&~~ \sum_{{ j,n,\ell
\in\bbz_+}} \mu_{j,n,\ell}~ \hm^\pm_\k ~\hp_{jn\ell} ~, &\opes a\cr 
T_\k ~\hp({\bar Y}) ~~=&~~ \sum_{{ j,n,\ell 
\in\bbz_+}} \mu_{j,n,\ell}~ T_\k ~\hp_{jn\ell} ~, &\opes b\cr }$$
where ~$\k ~=~ \x,\y,\z$, ~and the explicit action on
$\hp_{jn\ell}$ is defined by: \eqna\ops 
$$\eqalignno{
\hm^\pm_\x ~\hp_{jn\ell} ~=&~ \hp_{j\pm 1,n\ell} ~,
&\ops a\cr 
\hm^\pm_\y ~\hp_{jn\ell} ~=&~ \hp_{jn,\ell\pm 1} ~, 
&\ops b\cr 
\hm^\pm_\z ~\hp_{jn\ell} ~=&~ \hp_{j,n\pm 1,\ell} ~,
&\ops c\cr 
T_\x ~\hp_{jn\ell} ~=&~ q^{ j} ~\hp_{jn\ell} ~,
&\ops d\cr 
T_\y ~\hp_{jn\ell} ~=&~ q^{\ell} ~\hp_{jn\ell} ~, 
&\ops e\cr 
T_\z ~\hp_{jn\ell} ~=&~ q^{ n} ~\hp_{jn\ell} ~.
&\ops f\cr }$$ 

Now we define the $q$-difference operators by: 
\eqn\qdif{  \hd_\k ~\hp({\bar Y}) ~~=~~ {1\over \l} ~\hm^-_\k 
~\left( T_\k - T^{-1}_\k \right) ~ \hp({\bar Y}) ~, \quad   
\k ~=~ \x,\y,\z ~. }
Thus, we have: \eqna\opr 
$$\eqalignno{ 
\hd_\x ~\hp_{jn\ell} ~=&~ [j] ~\hp_{j-1,n\ell} ~, 
&\opr a\cr 
\hd_\y ~\hp_{jn\ell} ~=&~ [\ell] ~\hp_{jn,\ell-1} ~, 
&\opr b\cr 
\hd_\z ~\hp_{jn\ell} ~=&~ [n] ~\hp_{j,n-1,\ell} ~.
&\opr c\cr }$$
Of course, for ~$q\to 1$~ we have 
~$\hd_\k ~\to ~\pd_\k ~\equiv ~\pd/\pd \k$.

In terms of the above operators the transformation rules \abb{} are
written as follows: \eqna\abc 
$$\eqalignno{ 
\hat\pi_{r_1,r_2} (k_1) ~ \hp ({\bar Y}) ~~=&~~ q^{-r_1/2}  ~T_\x
~T_\z^{1/2} ~T_\y^{-1/2} ~\hp ({\bar Y}) ~, &\abc a\cr 
\hat\pi_{r_1,r_2} (k_2) ~ ~\hp ({\bar Y})
~~=&~~ q^{-r_2/2}  ~T_\y ~T_\z^{1/2} ~T_\x^{-1/2}  ~\hp ({\bar
Y}) ~, &\abc b\cr 
\hat\pi_{r_1,r_2} (X^+_1)  ~\hp ({\bar Y}) ~~=&~~ (1/\l) 
~\hm_\x ~T_\z^{1/2} ~T_\y^{-1/2}  ~\left( q^{-r_1} T_\x T_\z
T_\y^{-1} - q^{r_1} T_\x^{-1} T_\z^{-1} T_\y \right) 
 ~\hp ({\bar Y}) ~+&\cr &~~+~ 
q^{-r_1 -1} ~\hm_\z   ~\hd_\y ~T_\x ~T_\z^{1/2} ~T_\y^{-1/2} ~\hp
({\bar Y}) ~, &\abc c\cr 
\hat\pi_{r_1,r_2} (X^+_2)  ~\hp ({\bar Y}) 
~~=&~~(1/\l) ~\hm_\y ~T_\z^{1/2} ~T_\x^{-1/2}  ~\left( q^{-r_2} T_\y 
 - q^{r_2} T_\y^{-1}  \right)  ~\hp ({\bar Y}) ~-&\cr &~~-~ 
q^{r_2 -1} ~\hm_\z ~\hd_\x ~T_\x^{1/2} ~T_\z^{-1/2} ~T_\y^{-1} ~\hp
({\bar Y}) ~, &\abc d\cr 
\hat\pi_{r_1,r_2} (X^-_1)  ~\hp ({\bar Y}) 
~~=&~~ \hd_\x ~T_\z^{-1/2} ~T_\y^{1/2} ~\hp
({\bar Y}) ~, &\abc e\cr 
\hat\pi_{r_1,r_2} (X^-_2)  ~\hp ({\bar Y}) 
~~=&~~ -~ \hd_\y ~T_\z^{1/2} ~T_\x^{-1/2} ~\hp
({\bar Y}) ~-&\cr &~~-~ 
\hm_\x ~\hd_\z ~T_\x^{-1/2} ~T_\z^{1/2} ~T_\y^{-1} ~\hp
({\bar Y}) ~, &\abc f\cr } $$
where ~$\hm_\k ~=~ \hm^+_\k$~.

Notice that it is possible to obtain a realization of
the representation ~$\hat\pi_{r_1,r_2}$~ on monomials in three commuting
variables $x,y,z$. Indeed, one can relate the non-commuting
algebra ~$\bbc ~[\x,\y,\z]$~ with the commuting one ~$\bbc
~[x,y,z]$~ by fixing an ordering prescription. However, such
realization in commuting variables may be obtained much more
directly as is done by other methods and for other purposes in
\ref\BDT{L.C. Biedenharn, V.K. Dobrev and P. Truini, 
J. Math. Phys. {\bf 35} (1994) 6058.}. 
In the present paper we are interested in the non-commutative case
and we continue to work with the non-commuting variables
~$\x,\y,\z$.

\vskip 5mm 

Now we can illustrate some of the general statements of the
previous Section. Let ~$m_2 = r_2 +1 \in\bbn$. Then it is clear that
functions $\hp$ from \sersb{} with ~$\mu_{j,n,\ell} =0$~ if $\ell
\geq m_2$ form an invariant subspace since:
\eqn\sbs{\hat\pi_{r_1,r_2} (X^+_2)  ~\hp_{jnr_2} 
~~=~~ -~ q^{- 1 + (j - n )/2} ~[j]_q ~  ~\hp_{j-1,n+1,r_2} 
~, } 
and all other operators in \abb{} either preserve or lower the
index $\ell$. The same is true for the functions $\tv$. 
In particular, for ~$m_2=1$~ the functions 
in the invariant subspace do not depend on the variable ~$\y$. 
In this case we have functions of two $q$-commuting variables 
~$\z\x =q\x\z$~ which are much easier to handle that the general 
non-commutative case \coo. 

The intertwining operator \intc{} for ~$m_2 \in\bbn$~ 
is given as follows. First we calculate: 
\eqn\intd{ \eqalign{
\left( \pi_R(X^-_2)\right)^s ~\tv_{jn\ell} ~=&~ \left(
\pi_R(X^-_2)\right)^s ~ \x^j ~\z^n ~\y^\ell ~\cd^{r_1}_1 
\cd^{r_2}_2 ~= \cr 
=&~ \x^j ~\z^n ~\sum_{t=0}^s ~a_{st} ~\y^{\ell -t} ~\cd^{r_1
+t}_1  \cd^{r_2 -s-t }_2 ~(\x^{12}_{13})^{s-t} ~, \cr 
a_{st} ~=&~ q^{t\ell +r_2s/2 - (s+t)(s+t+1)/4} ~{s \choose t}_q ~
{ [r_2 -t]_q! [\ell]_q! \over [r_2 -s]_q! [\ell -t]_q! } ~,\cr}} 
where ~${n \choose k}_q ~\equiv ~[n]_q! /[k]_q! 
[n-k]_q!$, ~$[m]_q! ~\equiv ~[m]_q [m-1]_q \ldots [1]_q$. 
Thus, indeed $ \pi_R(X^-_2)$ is not preserving the reduced space
$\cc_{r_1,r_2}$, and  furthermore there is the additional variable
$\x^{12}_{13}$. Since we would like $ \pi_R(X^-_2)$ to some power
to map to another reduced space this is only possible if the
coefficients $a_{st}$ vanish for $s\neq t$. This happens iff 
~$s ~=~ r_2 + 1 ~=~ m_2$. Thus we have (in terms of the
representation parameters $m_i = r_i +1$):  
\eqn\inte{ \eqalign{
\left( \pi_R(X^-_2)\right)^{m_2} ~ \x^j ~\z^n ~\y^\ell ~&\cd^{m_1-1}_1 
\cd^{m_2-1}_2 ~=\cr  =&~ q^{ m_2 (\ell - 1- m_2/2)} ~
{ [\ell]_q! \over [\ell -m_2]_q! } ~\x^j ~\z^n ~\y^{\ell -m_2}
~\cd_1^{m_{12} -1} ~ \cd_2^{-m_2 -1} ~. \cr} }
Comparing the powers of $\cd_i$ we recover at once \redd\ for our
situation, namely, $m'_1 = m_{12}$, $m'_2 = -m_2$. Thus, we have
shown \op{a} and \inta{a}. Then \op{b} and \inta{b} follow using
\int. This intertwining operator has a kernel which is just the
invariant subspace discussed above - from the factor ~$1/  
[\ell -m_2]_q!$~ in \inte\ it is obvious that all
monomials with $\ell < m_2$ are mapped to zero. 

For the restricted functions we have: 
\eqn\intee{  \eqalign{ 
\left( \pi_R(X^-_2)\right)^{m_2} ~ \hp_{jn\ell} ~~=&~~ 
q^{ m_2 (\ell - 1- m_2/2)} ~ { [\ell]_q! \over [\ell -m_2]_q! } ~
\hp_{jn, \ell -m_2}  ~= \cr 
=&~~ q^{-3 m_2/2} ~\left( \hd_\y ~T_\y \right)^{m_2}
~ \hp_{jn\ell} ~.\cr }}
Thus, renormalizing \intc{b} by ~$q^{-3 m_2/2}$~ we 
finally have:
\eqn\intef{ I^{m_2}_2 ~~=~~ \left( \hd_\y ~T_\y \right)^{m_2}  ~. }
For $q=1$ this operator reduces to the known result:  
~$I_2 ~=~ ( \pd_\y )^{m_2}$ ~\Dob.

\vskip 5mm

Let now ~$m_1 \in\bbn$. 
In a similar way, though the calculations are more complicated,
we find: 
\eqn\intf{ \eqalign{
\left( \pi_R(X^-_1)\right)^{m_1} ~& \x^j ~\z^n ~\y^\ell ~\cd^{m_1-1}_1 
\cd^{m_2-1}_2 ~=\cr =&~~ q^{ m_1 (j+n-\ell - 1- m_1/2) }~
\sum_{t=0}^{m_1} ~q^{-t(t+3+2j)/2} ~\times \cr \times & ~{m_1
\choose t}_q ~
{ [j]_q! [n]_q!  \over [j -m_1 +t]_q! [n-t]_q! } ~
\x^{j+t-m_1} ~\z^{n-t} ~\y^{\ell +t} ~\cd_1^{-m_1 -1} ~
\cd_2^{m_{12} -1} ~. \cr} } 
Comparing the powers of $\cd_i$ we recover  \redd\ for our
situation, namely, $m'_1 = -m_1$, $m'_2 = m_{12}$. 
Thus, we have shown \op{} and \inta{}.

For the restricted functions we have: 
\eqn\intfe{  \eqalign{ 
\left( \pi_R(X^-_1)\right)^{m_1} ~ \hp_{jn\ell} ~~=&~~ 
q^{ m_1 (j+n-\ell - 1- m_1/2) }~
\sum_{t=0}^{m_1} ~q^{t(t+3+2j)/2} ~\times \cr & \times  ~{m_1
\choose t}_q ~
{ [j]_q! [n]_q!  \over [j -m_1 +t]_q! [n-t]_q! } ~
\hp_{j+t-m_1,n-t, \ell +t}  ~= \cr 
=&~~ q^{ -m_1 (3/2 + m_1) }~ T_\z^{m_1} ~ \sum_{t=0}^{m_1} 
~\hm_\y^t ~\hd_\z^t ~(q\hd_\x T_x)^{m_1-t} ~ T_\y^{-m_1} 
~\hp_{jn\ell} ~. \cr }} 
Then, renormalizing \intc{b} we finally have:
\eqn\intff{ I^{m_1}_1 ~~=~~ T_\z^{m_1} ~ \sum_{t=0}^{m_1} 
~\hm_\y^t ~\hd_\z^t ~(q\hd_\x T_x)^{m_1-t} ~ T_\y^{-m_1} ~.} 
For $q=1$ this operator reduces to the known result:  
~$I_1 ~=~ ( \pd_\x + \y  \pd_\z  )^{m_1}$ ~\Dob.

\vskip 5mm 

Finally, let us consider the case ~$m =m_{12} ~=~ m_1 + m_2
\in\bbn$, first with $m_1, m_2 \notin\bbn$. In this case the
intertwining operator is given by \intb, \sing\ with 
\ref\Doa{V.K. Dobrev, Talk at the International Group
Theory Conference (St. Andrews, 1989), Proceedings, Eds. C.M.
Campbell and E.F. Robertson, Vol. 1, London Math. Soc. Lecture
Note Series 159 (Cambridge University Press, 1991) pp. 87-104. 
}, 
formula (27), (cf. also \Doc):
\eqn\sng{ \eqalign{ 
\cp^m_{12} \left( X^-_1, X^-_2 \right) ~~=&~~ 
\sum_{s=0}^m ~a_s ~(X^-_1)^{m-s} ~(X^-_2)^m ~(X^-_1)^s ~, \cr 
a_s ~~=&~~ (-1)^s ~a ~{[m_1]_q\over [m_1 -s]_q}
~\left({m \atop s}\right)_q ~~, ~~s = 0,\dots,m , ~~a\neq 0 ~.
\cr} }

Let us illustrate the resulting intertwining operator in the case
$m=1$. Then, we have, setting in \sng\ ~$a = [1-m_1]_q$~: 
\eqn\snga{ \ci^{1}_{12} ~~=~~ [1-m_1]_q ~\pi_R (X^-_1)~ \pi_R (X^-_2)
~+~ [m_1]_q ~\pi_R (X^-_2)~ \pi_R (X^-_1) ~.} 
Then we can see at once the intertwining properties of 
$\ci^1_{12}$ by calculating: 
\eqn\sngb{ \eqalign{ 
\ci^1_{12} ~\x^j ~\z^n ~\y^\ell ~\cd^{m_1-1}_1 
\cd^{m_2-1}_2 ~~=&~~ q^{j+n-2-m_1} ~[j]_q ~[\ell]_q ~\x^{j-1} ~\z^n
~\y^{\ell -1} ~\cd^{m_1-2}_1  \cd^{m_2-2}_2 ~+\cr 
&+~ q^{n-2} ~[n]_q ~[\ell + m_1]_q ~\x^j ~\z^{n-1}
~\y^\ell ~\cd^{m_1-2}_1  \cd^{m_2-2}_2 ~. \cr } }
Comparing the powers of $\cd_i$ we recover \redd\ for our
situation, namely, $m'_1 = -m_2 = m_1 -1$, $m'_2 = -m_1 =m_2 -1$. 

For the restricted functions we have: 
\eqn\sngc{ \eqalign{ 
\bigl( [1-m_1]_q& ~\pi_R (X^-_1)~ \pi_R (X^-_2)
~+~ [m_1]_q ~\pi_R (X^-_2)~ \pi_R (X^-_1) \bigr) 
 ~ \hp_{jn\ell} ~~=\cr 
=&~~ q^{n-2 +j-m_1} ~[j]_q ~[\ell]_q ~ \hp_{j-1,n,\ell-1}
~+~ q^{n-2} ~[n]_q ~[\ell + m_1]_q ~ \hp_{j,n-1,\ell} ~=\cr 
=&~~ q^{-2} ~\left( q^{-m_1}~ \hd_\x ~T_\x ~ \hd_\y ~+
~ (1/\l) ~\hd_\z ~( q^{m_1} T_\y - q^{-m_1} T_\y^{-1}) 
\right) ~T_\z   ~ \hp_{jn\ell} ~. \cr } } 
Rescaling \intb{b} we finally have: 
\eqn\sngd{ I^{1}_{12} ~~=~~ \left( q^{-m_1}~ \hd_\x ~T_\x ~ \hd_\y ~+
~ (1/\l) ~\hd_\z ~( q^{m_1} T_\y - q^{-m_1} T_\y^{-1}) 
\right) ~T_\z  ~. } 
For $q=1$ this operator is: 
~$I_{12} ~=~  \pd_\x \pd_\y + (m_1+ \y\pd_\y) \pd_\z$ ~ 
\Dob. 

Above we have supposed that $m_1, m_2\notin\bbn$. However, after
the proper choice of $a$ in \sng, (e.g., as made above in \snga) 
we can consider the singular vector \sng\ and the resulting 
intertwining operator also when $m_1$ and/or $m_2$ are 
positive integers. Of particular interest are the 
cases $m_1,m_2\in\bbz_+$. In these cases the singular vector 
is reduced in four different ways (cf. \Doa, \Doc\ formulae (33a-d)).
Accordingly, the intertwining operator becomes composite, 
i.e., it can be expressed as the composition of the 
intertwiners introduced so far as follows: 
\eqna\snge 
$$\eqalignno{ I^m_{12} ~&=~ c_1~ I^{m_2}_1  ~I^{m}_2  ~I^{m_1}_1 ~=   
&\snge a\cr 
&=~ c_2~ I^{m_1}_2  ~I^{m}_1  ~I^{m_2}_2   ~= 
&\snge b\cr 
&=~ c_3 ~ I^{m_1}_2  ~I^{m_2}_{12}  ~I^{m_1}_1   ~= 
&\snge c\cr 
&=~ c_4 ~ I^{m_2}_1  ~I^{m_1}_{12}  ~I^{m_2}_2   ~.  
&\snge d\cr }$$ 
The four expressions were used to prove commutativity 
of the hexagon diagram of $U_q(sl(3,\bbc))$) \Doa. 
This diagram involves six representations which are 
denoted by $V_{00}$, $V^1_{00}$, $V^2_{00}$, $V^{12}_{00}$,  
$V^{21}_{00}$,  $V^3_{00}$, in (29) of \Doa\ and which in 
our notation are connected by the intertwiners in \snge{} as follows: 
\eqna\sngf 
$$\eqalignno{ 
&\tc_{m_1,m_2} \quad {I^{m_1}_1 \atop \rra} 
\quad  \tc_{-m_1,m} \quad {I^{m}_2 \atop \rra} 
\quad  \tc_{m_2,-m} \quad {I^{m_2}_1 \atop \rra} 
\quad \tc_{-m_2,-m_1} ~, &\sngf a\cr &&\cr 
&\tc_{m_1,m_2} \quad {I^{m_2}_2 \atop \rra} 
\quad  \tc_{m,-m_2} \quad {I^{m}_1 \atop \rra} 
\quad  \tc_{-m,m_1} \quad {I^{m_1}_2 \atop \rra} 
\quad \tc_{-m_2,-m_1} ~, &\sngf b\cr &&\cr 
&\tc_{m_1,m_2} \quad {I^{m_1}_1 \atop \rra} 
\quad  \tc_{-m_1,m} \quad {I^{m_2}_{12} \atop \rra} 
\quad  \tc_{-m,m_1} \quad {I^{m_1}_2 \atop \rra} 
\quad \tc_{-m_2,-m_1} ~, &\sngf c\cr &&\cr 
&\tc_{m_1,m_2} \quad {I^{m_2}_2 \atop \rra} 
\quad  \tc_{m,-m_2} \quad {I^{m_1}_{12} \atop \rra} 
\quad  \tc_{m_2,-m} \quad {I^{m_2}_1 \atop \rra} 
\quad \tc_{-m_2,-m_1} ~. &\sngf d\cr 
}$$ 
Of these six representations only $\tc_{m_1,m_2}$ 
has a finite dimensional irreducible subspace 
iff $m_1m_2>0$, the dimension being $m_1 m_2 m/2$ \Doa.  
If $m_1=0$ the intertwining operators with superscript $m_1$ 
become the identity (since in these cases the 
intertwined spaces coincide) and the compositions in \snge{}, 
\sngf{} are shortened to two terms in cases (a,b,d) and one term 
in case (c), (resp. for $m_2=0$, two terms in cases (a,b,c), one 
term in (d)). (Such considerations are part of the multiplet 
classification given in \Doa.)

\vskip 1cm 

\noindent 
{\bf Note Added published in: J. Phys. A: Math. Gen. {\bf 27}
(1994) 6633-6634.}  

\noindent 
{\bf 1.} ~ We decided to add several formulae which will be useful 
for those who would like to consider in more detail $U_q(sl(n))$ 
for $n>3$ without waiting for the sequel of this article. 
(Some of these formulae were used above in their simple $n=3$ 
versions without explication.) First 
we give the commutation relations of the $Y_{j\ell}$ and $D_i$ 
variables: \eqna\cop \eqna\cor  
$$\eqalignno{ 
Y_{i\ell} Y_{ij} ~&=~ q Y_{ij} Y_{i\ell}   ~, 
~~ i> \ell >j ~, &\cop a\cr  
Y_{kj} Y_{ij} ~&=~ q Y_{ij} Y_{kj}   ~, ~~k> i > j ~, &\cop
b \cr 
Y_{i\ell} Y_{kj} ~&=~ Y_{kj} Y_{i\ell} ~, ~~ k>i> \ell>j ~, &\cop c \cr  
Y_{ij} Y_{k\ell} ~&=~ Y_{k\ell} Y_{ij} -    \l
Y_{i\ell} Y_{kj} ~,  ~~ k>i,\ell>j  ~, ~~i \neq \ell ~, 
&\cop d \cr
Y_{ij} Y_{ki} ~&=~ qY_{ki} Y_{ij} -  \l  Y_{kj} ~,  ~~ k>i >j  ~,
&\cop e \cr 
Y_{j\ell} D_{i} ~&=~  D_{i} Y_{j\ell}   ~, 
~~ j> \ell > i ~, &\cor a\cr 
Y_{j\ell} D_{i} ~&=~ q D_{i} Y_{j\ell}   ~, 
~~ j> i \geq \ell ~, &\cor b\cr 
Y_{j\ell} D_{i} ~&=~  D_{i} Y_{j\ell}   ~, 
~~ i \geq j >\ell ~, &\cor c\cr 
}$$ 
where in \cop{d} we use ~$Y_{i\ell} =0$~ when ~$i<\ell$. Note
that \cop{a-d} may be obtained by replacing ~$a_{i\ell}$~
with ~$Y_{i\ell}$~ in \1{a-d}. Note that the structure of the
$q$ - flag manifold for general ~$n$~ is exibited already for
~$n=4$, while for ~$n=3$~ relations \cop{c,d} are not present 
- cf. \coo. The commutation relations between the $Z$ and $D$ 
variables are obtained from \cop{}, \cor{}, by just replacing 
$Y_{st}$ by $Z_{ts}$ in all formuale. \hfil\break 
Next, we explicate the right action on the variables $Z_{j\ell}$: 
\eqna\rcfff
$$\eqalignno{ \pi_R (X^+_i) ~Z_{j\ell} ~=&~~ \d_{i+1,\ell}~ 
q^{\d_{ij}/2}~ Z_{j,\ell-1} ~, &\rcfff a\cr 
\pi_R (X^-_i) ~Z_{j\ell} ~=&~ \d_{i\ell} ~Z_{j,\ell+1} 
~-~ \d_{ij}~ q^{-\d_{j+1,\ell}/2} ~Z_{j,j+1} ~Z_{j\ell} ~
+~  \d_{i,j-1} ~ D_j^{-1} ~\xiL{j}{j-2,j,\ell} &\rcfff b\cr 
\pi_R (k_i) ~Z_{j\ell} ~=&~ q^{(
\d_{i+1,j} - \d_{ij} + \d_{i\ell} - \d_{i+1,\ell})/2} ~
Z_{j\ell} ~. &\rcfff c\cr}$$  
Formula \rcfff{a} may have appeared after \rcff, while 
the other two  are used in the calculation of the 
intertwiners. In the latter calculations we also use: 
\eqna\lar 
$$\eqalignno{
\pi_R (X^-_i) ~(D_\ell)^n 
~~&=~~ \d_{i\ell} ~c_n ~(D_\ell)^n ~Z_{\ell,\ell+1} ~~, 
&\lar a\cr 
\pi_R (X^-_i) ~(Y_{j\ell})^n 
~~&=~~ \d_{i\ell} ~q^{n-3/2}~[n]_q ~(Y_{j\ell})^{n-1}~ 
Y_{j,\ell+1} ~D_{\ell +1} ~ D_{\ell}^{-2} ~ D_{\ell -1} ~. 
&\lar b\cr}$$    
\hfil\break \noindent 
{\bf 2.} ~A $q$ - difference operator realization of $U_q(sl(3))$ 
depending on two $q$-commuting variables and one 
integer representation 
parameter was constructed by a different method 
in \ref\FV{R. Floreanini and L. Vinet, 
Phys. Lett. {\bf 315B} (1993) 299.}. Thus, formula (24) of \FV\ 
should be compared with our \abc{} if we set in \abc{} $r_1\in\bbz_+$, 
$r_2=0$, $T_\y = $~id, $\hd_\y =0$, and then restrict our functions 
to the variables $\x,\z$. 

\vfil\eject 

\baselineskip=14pt 

\listrefs 

\vfil\eject \end